\documentclass[twocolumn,amsmath,amssymb,aps,prl]{revtex4}
\usepackage{amsmath,graphicx,dcolumn,bm}

\begin{document}

\title{Alpha Channeling in a Rotating Plasma}

\author{Abraham J. Fetterman}
\author{Nathaniel J. Fisch}
\affiliation{
Department of Astrophysical Sciences, 
Princeton University, 
Princeton, New Jersey 08540, USA}

\date{\today}

\begin{abstract}
  The wave-particle $\alpha$-channeling effect is generalized to
  include rotating plasma.  Specifically, radio frequency waves can
  resonate with $\alpha$ particles in a mirror machine with $E \times
  B$ rotation to diffuse the $\alpha$ particles along constrained
  paths in phase space.  Of major interest is that the
  $\alpha$-particle energy, in addition to amplifying the RF waves,
  can directly enhance the rotation energy which in turn provides
  additional plasma confinement in centrifugal fusion reactors.  An
  ancillary benefit is the rapid removal of alpha particles, which
  increases the fusion reactivity.
\end{abstract}

\maketitle

In magnetic mirror fusion devices, centrifugal forces can
significantly enhance the magnetic confinement~\cite{Lehnert:1971p527,
  Bekhtenev1980, Pastukhov1987}.  A radial electric field induces
rapid $\mathbf{E}\times\mathbf{B}$ plasma rotation, leading to the
centrifugal force that directly confines ions axially.  Electrons are
then confined axially through the ambipolar potential. The radial
field thus not only enhances the plasma confinement, but also produces
the necessary heating for the plasma, as injected cold neutral fuel
atoms are seen as moving at the rotation velocity in the rotating
frame.  Lately there has been a renewed interest in this
effect~\cite{Bekhtenev1977, Abdrashitov1991, Hassam1992,
  Ellis:2000p95}, strengthened by recent findings of reduced
turbulence due to sheared rotation~\cite{Burrell:1997p844,
  Cho:2005p376}.

What we show here is that in a DT (deuterium-tritium) centrifugal
fusion reactor, the energy of $\alpha$-particles, the byproducts of
the fusion reaction, might be advantageously induced to directly
produce this rotation.  The predicted effect relies on exploiting the
population inversion of the birth distribution of $\alpha$
particles. This is a generalization of the {\it alpha channeling
  effect}, where injected wave energy can be amplified at the expense
of the $\alpha$-particle energy, with the alpha particles
concomitantly removed as cold particles \cite{Fisch:1992p78}.  In
tokamaks, if the wave energy is damped on ions, the fusion reactivity
might be doubled \cite{Fisch1994}.  Similar advantageous uses of
$\alpha$-channeling can be expected in mirror machines
\cite{fisch:225001}.  With several waves, a significant amount of the
$\alpha$-particle energy can be advantageously channeled in both
tokamaks \cite{Herrmann:1997p60} and mirrors \cite{Zhmoginov2008}.
However, in previous considerations of $\alpha$-channeling, the plasma
was not rotating strongly.

In strongly rotating plasma, significant new effects can occur because
there are two further reservoirs of particle energy, namely rotational
and potential energy.  For example, through a suitable choice of wave
parameters, particles can now absorb wave energy yet cool in kinetic
energy, with the excess energy being stored in potential
energy. Alternatively, particle potential energy might be lost to wave
energy with kinetic energy constant.  These possibilities could not be
achieved through particle manipulation in stationary systems, where
the only coupling is between the kinetic energy with the wave energy.
What is important for centrifugal mirror fusion is that radiofrequency
waves can drive a radial $\alpha$-particle current, with the
dissipated power extracted from the $\alpha$-particle birth energy,
thereby maintaining the radial potential which produces the necessary
plasma rotation.

To derive the new effects, define the angular rotation frequency
$\mathbf{\Omega_E}=\Omega_E\hat{z}$, so that the
$\mathbf{E}\times\mathbf{B}$ drift velocity can be written as
$\mathbf{\Omega}_E\times\mathbf{r} = \mathbf{E}\times \mathbf{B}/B^2$.
For simplicity, consider constant $\Omega_E$ (solid-body rotation).
Although some aspects may vary with the rotation profile, the concept
should be applicable to arbitrary profiles $\Omega \left( r
\right)$. The electric and magnetic field in the rotating frame
are~\cite{Lehnert1964},
\begin{align}\label{eq:Etilde}
\tilde{\mathbf{E}}& =  \mathbf{E}+\frac{m}{q}\Omega^2\mathbf{r}+
\left(\mathbf{\Omega}\times\mathbf{r}\right)\times\mathbf{B}, \\
\tilde{\mathbf{B}} & =\mathbf{B}+2\frac{m}{q}\mathbf{\Omega}.\label{eq:Btilde}
\end{align}
The second term in Eq.~(\ref{eq:Etilde}) produces the centrifugal
force, and the second term in Eq.~(\ref{eq:Btilde}) is due to the
Coriolis effect.  For $\mathbf{\Omega}=\mathbf{\Omega}_E$, the first
and third terms in Eq.~(\ref{eq:Etilde}) will cancel.  However, there
will still be drifts due to the centrifugal force.  We define as
$\Omega_E^\star$ the unique frame of reference in which
$\hat{\theta}\cdot \tilde{\mathbf{E}} \times \tilde{\mathbf{B}}=0$ for
the species of interest.  Note that the magnetic moment is seen as
invariant only in the frame rotating with frequency $\Omega_E^\star$.
Our notational convention is to denote terms in this frame with a
tilde.

\begin{figure}[b]
\centering
\includegraphics[scale=.4]{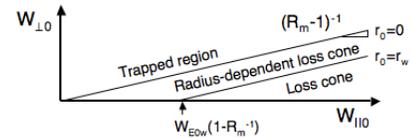}
\caption{\label{fig:losscone}The loss cone in (rotating) midplane
  energy coordinates for a rotating plasma, including centrifugal
  confinement.}
  %
  %  Denote W_par with a tilde even though it is unchanged in the rotating frame
  %
\end{figure}

Note that by flux conservation, $r^2/r_0^2\propto
\tilde{B}/\tilde{B}_0$.  Thus for magnetic mirror ratio
$R_m=\tilde{B}_m/\tilde{B}_0$, there is an effective confinement
potential $\Phi_c=\frac{1}{2} m\Omega_E^{\star 2}r_0^2
\left(1-R_m^{-1}\right)$, which varies with the midplane particle
radius $r_0$.  The loss-cone diagram is depicted in
Fig.~\ref{fig:losscone}.  The maximum confinement potential is
$W_{E0w}\left(1-R_m^{-1}\right)$, where $W_{E0w}=\frac{1}{2}
m\Omega_E^{\star 2}r_w^2$, and $r_w$ is the midplane radius of the
last field line not intersecting a wall.

\begin{figure}
\centering
\includegraphics[scale=.4]{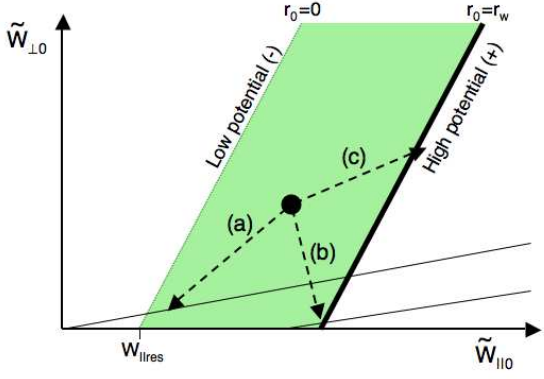}
\caption{\label{fig:diffpaths}The shaded region is the region of wave resonance, dependent on radius, for interaction with an RF wave  at mirror ratio $R_{rf}$.  The hatched lines depict three diffusion paths that would
eject particles, using only perpendicular diffusion. Path
(a) reduces both the kinetic and potential energy of the particle,
path (b) increases potential energy but decreases kinetic energy,
and path (c) increases both kinetic and potential energy. The energy
balance is assumed by the wave.}
\end{figure}

Now consider a wave with frequency $\omega$, parallel wave number
$k_\|$ and azimuthal mode number $n_\theta=k_\theta r$.  Due to the
rotation, the wave frequency in the rotating frame will be
$\tilde{\omega}=\omega-n_\theta \Omega_E^\star$.  The wave-particle
resonance condition is then
$\tilde{\omega}-k_\|v_\|=n\tilde{\Omega}_c$, where the resonance is at
the $n^\mathrm{th}$ harmonic of the rotating-frame cyclotron frequency
$\tilde{\Omega}_c=q\tilde{B}/m$ ($q$ is the ion charge and $m$ is its
mass).  The parallel velocity $v_\|$ is independent of the rotating
frame, and corresponds to energy $W_{\| res}=mv_\|^2/2$.  Unlike in
the stationary case, the related midplane parallel energy, $W_{\|
  0res}$, will not be constant across the radius of the device. For an
RF region at mirror ratio $R_{rf}=\tilde{B}_{rf}/\tilde{B}_0$, the
resonant parallel energy in rotating midplane coordinates is
\begin{equation}\label{Wll0res}
W_{\|  0res} = W_{\| res}+W_{E0}\left(1-R_{rf}^{-1}\right),
\end{equation}
where $W_{E0}=m\Omega_E^{\star 2}r_0^2/2$.  These
resonant regions appear as the shaded region in Fig.~\ref{fig:diffpaths}.

Since the mirror system is axisymmetric, the diffusion paths will be the
same as those for tokamaks~\cite{Fisch:1992p78},
\begin{align}\label{dPthetadW}
dP_\theta/d\tilde{W}& =n_\theta/\tilde{\omega},\\
d\tilde{\mu}/d\tilde{W} & =qn/m\tilde{\omega},\label{dmudW}
\end{align}
where $\tilde{\mu}=m\tilde{v}_\perp^2/2\tilde{B}$ is the ion magnetic
moment in the rotating frame;
$\tilde{W}=\tilde{\mu}\tilde{B}+mv_\|^2/2$ is the kinetic energy in
the rotating frame; and $P_\theta$ is the azimuthal canonical angular
momentum (which is frame-independent).  

The significant difference in the rotating frame is that the
interaction of the particle with a wave at axial position $z_{rf}$
changes the particle's perpendicular, parallel, and rotational kinetic
energy, as well as its potential energy.  The change in perpendicular
energy may be written $\tilde{W}_\perp \left(z_{rf} \right)
\rightarrow \tilde{W}_\perp \left(z_{rf}\right)+\Delta
\tilde{W}_\perp$; the change in parallel energy, $\tilde{W}_\|
\left(z_{rf}\right) \rightarrow \tilde{W}_\| \left( z_{rf}\right) +
\Delta \tilde{W}_\|$; and the change in rotational energy,
$W_E\left(z_{rf} \right) \rightarrow W_E\left(z_{rf}\right) +\Delta
W_E$.  Thus the wave interaction, breaking the adiabatic invariance of
$\tilde{\mu}$, gives stochastic kicks in $\Delta \tilde{W}_\perp$,
$\Delta W_\|$, and $\Delta W_E$.

The energy kicks are correlated through the properties of the wave.
The relation between $\Delta\tilde{W}_\perp$ and $\Delta \tilde{W}_\|$
is found, by Eq.~(\ref{dmudW}), to be $\Delta \tilde{W}_\|= \Delta
\tilde{W}_\perp k_\|v_\| /(n\tilde{\Omega}_c)$.  The radial excursion
is determined in terms of the perpendicular energy change by
Eq.~(\ref{dPthetadW}), yielding $r\Delta r= \Delta \tilde{W}_\perp
n_\theta/ (m\tilde{\omega}\tilde{\Omega}_c)$.  This then gives the
rotational energy change, $\Delta W_E=m\Omega_E^{\star 2}r\Delta
r=\Delta \tilde{W}_\perp n_\theta \Omega_E^{\star 2}
/(\tilde{\omega}\tilde{\Omega}_c)$, and the potential energy change,
$q\Delta \Phi = -qE\Delta r = n_\theta \Omega_E \Omega_c /
(\tilde{\omega} \tilde{\Omega}_c) \Delta \tilde{W}_\perp$.

Using the adiabatic invariance of $\tilde{\mu}$, flux conservation
($r^2/r_0^2\propto \tilde{B}/\tilde{B}_0$), and conservation of
energy, we require
\begin{equation}\label{wllenergycons}
  \Delta \tilde{W}_\perp + \Delta \tilde{W}_\| - \Delta W_E =
  R_{rf}^{-1} \Delta \tilde{W}_\perp + \Delta \tilde{W}_{\| 0} - R_{rf} \Delta W_E ,
\end{equation}
so that the changes in rotating midplane coordinates can be written as,
\begin{align}
  \label{DWperp0} \Delta\tilde{W}_{\perp 0}& = \Delta \tilde{W}_\perp/R_{rf},\\
  \label{DWll0} \Delta \tilde{W}_{\| 0} & = \left[\frac{k_\|
      v_\|}{n\tilde{\Omega}_c} + \left(R_{rf}-1\right)
    \frac{n_\theta\Omega_E^{\star 2}}{\tilde{\omega}\tilde{\Omega}_c}
    + \left(1-R_{rf}^{-1}\right)\right]\Delta \tilde{W}_\perp , \\
  \label{Dr0} \Delta r_0 & =\frac{R_{rf}n_\theta}
  {mr_0\tilde{\omega}\tilde{\Omega}_c}\Delta \tilde{W}_\perp .
\end{align}

As the particle diffuses in radius it also changes its rotation energy
$W_E$.  This will lead to a change in midplane parallel energy for
$R_{rf}>1$, as can be seen in Eq.~(\ref{Wll0res}).  This is the source
of the second term in brackets in Eq.~(\ref{DWll0}).  Note that the
particle remains in resonance with the wave on its entire path in the
limit $n_\|\rightarrow 0$.

With reference now to Fig.~\ref{fig:diffpaths}, note three ways
particles might be extracted from a rotating mirror, with 
perpendicular diffusion only ($\Delta \tilde{W}_{\|} =0$). The particle begins midway between the axis
and wall of the device. Suppose the particle may be removed through the loss
cone by path (a) at a low potential energy and a low kinetic
energy. This requires the wave phase velocity in the rotating frame to
be positive ($\tilde{v}_{p}=k_{\theta}/\tilde{\omega}>0$).  The same
wave may be used to remove particles through the last flux surface at
high kinetic and potential energy (shown by path (c)).  The energy
balance in each case is carried by the interacting wave.  Path (b)
describes a diffusion path where the particle is removed with less
kinetic energy than at its birth but at a higher potential energy. The
particle may be removed either through the loss cone or the last flux
surface.  This is the useful case for maintaining the radial electric
field.

To calculate the {\it branching ratio} $f_{E}$, the ratio of energy
going into the radial potential to the total energy change, consider
that the change in rest-frame kinetic energy is
\begin{equation}
  \Delta W=  \left(\frac{\omega}{\tilde{\omega}} + \frac{k_\| 
      v_\|}{n\tilde{\Omega}_c}\right) \Delta \tilde{W}_\perp,  
\end{equation}
giving the branching ratio
\begin{align}\label{fE}
f_E & = 
    \frac{-n_\theta \Omega_E
\Omega_c}{\omega \tilde{\Omega}_c + \tilde{\omega}k_\| v_\| /n
-n_\theta \Omega_E \Omega_c}, \\
 &\approx  \label{fEapprox}
\frac{-n_\theta \Omega_E}{\Omega_c + 2 k_\| v_\| + 4 \Omega_E} ,
\end{align}
where the approximation in Eq.~(\ref{fEapprox}) uses the resonance
condition $\tilde{\omega}=\tilde{\Omega}_c+k_\| v_\|$, with
$\Omega_E,k_\| v_\| \ll \Omega_c$.  If these conditions are
sufficiently strong, the fraction of the total energy change provided
to the radial electric field is $f_E\approx
-n_\theta\Omega_E/\Omega_c$.  In the case $f_E>1$, the particle
reduces its kinetic energy and simultaneously absorbs wave energy,
which can be expected because the direction of the RF wave
phase velocity in the rotating frame,
$\tilde{v}_p=\tilde{\omega}/k_\theta$ is opposite that in the
laboratory frame, $v_p=\omega/k_\theta$.  Path (b) in
Fig.~\ref{fig:diffpaths} describes a diffusion path in which the wave
will be amplified if $f_{E}<1$ (not all kinetic energy is converted to
potential), or damped if $f_{E}>1$ (wave energy transferred to
potential energy).

The waves necessary for a channeling effect have been calculated in
the static mirror case~\cite{fisch:225001,Zhmoginov2008}.  Two
conditions were considered to be important.  In order for the
diffusion path to be favorable, it must connect a dense area of
phase-space near the birth population to a less dense area of phase
space near the loss boundary.  In addition, it is advantageous that
the $\alpha$ particle heating be limited above the birth energy.
It was shown that waves with purely perpendicular diffusion ($\Delta
W_\|=0$, $n=1$, $k_\|v_\| \ll \tilde{\Omega}_c$) and $T_i \ll W_{\|
  res} \ll W_{\alpha 0}$ satisfy the first requirement.  In the
rotating system, the results are the same: perpendicular diffusion
with $W_{E 0} < W_{\| res} \ll W_{\alpha 0}$ will provide connection
to the velocity-space loss cone, and $\alpha$ particles will leave at
low energy.

The limitation of the $\alpha$ particle heating along the diffusion
path may be accomplished in two ways~\cite{Zhmoginov2008}.  The first
way is by diffusing energy-gaining particles inward, since the
particle is restricted to $r\geq 0$.  This is a strong limit to the
energy gain, but requires large $k_\theta$ to limit the energy gain to
a few MeV for large devices (see Eq.~(\ref{Dr0})).  The second way to
limit $\alpha$-particle heating is by noting that the diffusion
coefficient of a particle in a wave near harmonic $n$ of the cyclotron
frequency is proportional to $J_n^2(k_\theta \rho)$, where $J_n$ is
the Bessel function of the first kind and $\rho$ is the
gyroradius~\cite{Karney:1979p853}.  The diffusion coefficient will be
nearly zero if $k_\theta \rho$ is equal to a zero of the Bessel
function.  For the first Bessel zero to be above the $\alpha$-particle
birth energy, $k_\theta < 3.8317/\rho_{\alpha 0}$ is required. 

In any case, it will improve efficiency to have particles gaining
energy move inward, where they are less likely to be lost at high
energy.  Then by Eq.~(\ref{Dr0}), $n_\theta$ must be
negative.  For an outward current to maintain the potential, $\Omega_E$
must be positive. This is also the preferred polarity for rotating mirror
systems~\cite{Lehnert:1971p527}.

Efficient conversion of alpha birth energy to potential energy can be
achieved by selecting $n_\theta$ such that $f_E \sim 1$.  For
$B_0=1\,\mathrm{T}$, this means that $-n_\theta \Omega_E \sim 50\times
10^{6}\, \mathrm{s^{-1}}$.  An electric field of $50\,\mathrm{kV/cm}$
at $r=1\,\mathrm{m}$ produces $\Omega_E=5\times 10^6/\mathrm{s}$.  To
convert all of the $\alpha$ particle kinetic energy to electric
potential energy, choose $n_\theta=10$, or at $r=1\,\mathrm{m}$,
$k_\theta\approx 0.1/\mathrm{cm}$.  This is also below the first
Bessel function zero, which for 3.5 MeV $\alpha$ particles is near
$k_\theta=0.14/\mathrm{cm}$.

\begin{figure}
\centering
\includegraphics[scale=0.68]{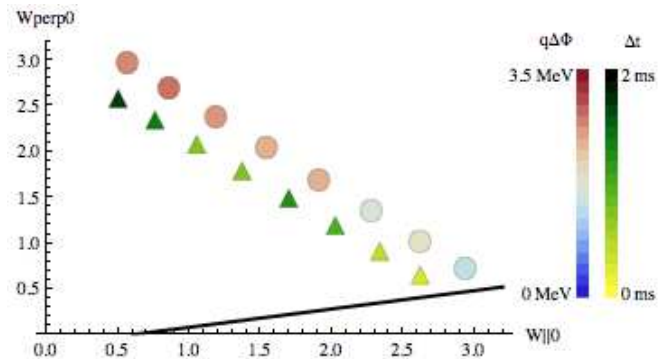}
\caption{\label{fig:compdata} Numerical simulation of alpha
  channeling. The circles indicate the amount of energy channeled into
  the radial potential. The triangles indicate the loss time. Both
  circles and triangles refer to the same set of alpha particles born
  at 3.5 MeV; their position indicates the ratio $W_{\perp0}/W_{\|0}$
  at birth. Each point is the average of 20 single-particle
  simulations. }
\end{figure}

We simulated this alpha channeling effect in a rotating mirror plasma, by following the full equations of motion for 160 particles interacting with ten RF
regions.  The wave parameters $W_{\|res}=750\,\mathrm{keV}$,
$k_{\|}=-0.03/\mathrm{cm}$, $E_{rf}=30\,\mathrm{kV/cm}$ were constant
across each wave, but $n_{\theta}=12-20$ and $R_{rf}=1.2-4.5$
varied. The mirror ratio for the simulation was $R_{m}=6$, and the
device length was 20 m.  The statistics were sufficient to
estimate that 65\% of the energy was channeled into the radial
potential.  Fig.~\ref{fig:compdata} depicts, for these parameters, the
relative effectiveness of energy extraction from alpha particles of
different birth energies, as well as the relative time for extraction.
The energy extraction needs to be completed before the collisional
slowing down of the alpha particle.

The primary application of the $\alpha$ channeling effect proposed
here is to use the energy of $\alpha$ particles to support the
rotation of the plasma, which is the main power requirement in
centrifugal fusion reactors; the heating of the fuel is then automatic
since through ionization new fuel particles are born with a high
kinetic energy in the rotating frame.  Define the fusion energy gain,
$Q \equiv P_{\rm f}/P_{\rm circ}$, as the ratio of fusion power to
circulating power to maintain the rotation.  Let $\eta$ be the
fraction of alpha particle power $P_{\rm f}/5$ that supports the
rotation, then the fusion energy gain in the presence of the $\alpha$
channeling effect can be written as $Q_{\rm AC} = Q/(1- \eta
Q/5)$. Thus, if all of the $\alpha$ particle energy could be converted
into rotation energy ($\eta=1$), a reactor formerly operating at $Q=5$
would become self-sustaining, requiring no external heating or energy
input.

There are several considerations to address in choosing the branching
ratio $f_E$ in each wave region.  In order to extract efficiently the
energy of the alpha particles, they must be removed by diffusion
(having rate proportional to the standing wave energy) before they are
lost collisionally.  Thus, the branching ratio should be set first to
assure that the waves reach sufficient amplitude for collisionless
diffusion.  On the other hand, it is not beneficial to amplify the
wave beyond what is needed to satisfy this criterion.  The power going
into the waves in other fusion devices is generally thought to be most
effectively channeled into fuel ion heating
\cite{Fisch:1992p78,fisch:225001}; in contrast, the further heating of
ions is not necessary in rotating mirrors, since they are born at high
energy.  To the extent that extra power is used to support the
rotation, beyond the necessary rotation for confinement, the reactor
essentially acts as a battery, producing an EMF source that may be
loaded through the end electrodes in the same way as a hydromagnetic
capacitor~\cite{Anderson1959}.  Most likely, the optimum design would
just maintain the rotation and the necessary diffusion time.

Many implementations of open systems attempt the direct conversion of
charged fusion product energy into electrical
energy~\cite{Post1987,Volosov1997}.  A recent suggestion for
centrifugal confinement devices \cite{Volosov2006} captures $\alpha$
particle energy both through the centrifugal potential (which goes
directly into the rotation energy) and through a retarding potential
(direct conversion).  But the amount of energy extractable to maintain
the rotation energy is a small fraction of the alpha particle kinetic
energy.  In contrast, as proposed here, the $\alpha$ particle energy
can be almost entirely converted into potential energy.

The fact that the effect proposed here acts volumetrically -- not at a
surface -- may importantly alleviate the major engineering hurdle
facing the use of rotating mirrors as fusion
reactors~\cite{Lehnert:1971p527,Bekhtenev1980, Abdrashitov1991},
namely the endplate electrodes.  These electrodes need to support
large electric fields which are subject to breakdown.

Because centrifugal fusion reactors are run in the hot ion mode ($T_i
> T_e$), the alpha channeling effect is particularly fitting.  The
ions are hot in rotating plasma because they are born at the rotation
speed.  The rapid removal of $\alpha$ particles, which are slowed down
primarily by electrons, then removes an important electron heat
source, thereby permitting an even cooler electron temperature.  The
cooler electron temperature in turn gives rise to a lower ambipolar
potential, which means higher ion confinement.  Higher ion
confinement, in turn, then relaxes the need for additional rotational
confinement, so that the mirror can be operated at lower rotation
speeds and lower plasma potential.  In addition to reducing the
electron heating, the quick expulsion of alpha particles by waves
reduces the dilution of fuel ions by the alpha particles. Like in a
conventional mirror reactor, where the alpha particle ash can dilute
the fuel by as much as 30\% \cite{Fowler1966}, the prompt removal of
this ash (and the channeling of that energy to fuel ions) can increase
greatly the effective fusion reactivity at fixed plasma pressure
\cite{fisch:225001}.

In conclusion, we generalized the alpha channeling effect to rotating
plasma.  A new quantity that appears in rotating plasma is the
branching ratio, which measures the amount of particle kinetic energy
that flows into particle potential energy as compared to the amount
which flows into wave energy.  By arranging for sufficient channeling
of fusion alpha energy directly into electric potential energy, the
rotation of the plasma can be maintained against momentum loss. The
prompt removal of alpha particles also increases the effective fusion
reactivity.  Also, the volumetric maintenance of the radial potential
should  reduce the engineering complexity of the
technologically challenging mirror endplates, if not to eliminate the
need for these plates entirely.  Moreover, the alpha channeling effect
is particularly well matched to enhance the reactor prospects of
centrifugal fusion reactors, since these reactors are imagined to
operate best at low electron temperature and high ion temperature.
While the channeling of alpha energy in rotating plasma appears to
significantly enhance the prospects for controlled nuclear fusion
through centrifugal confinement, it
remains to identify the specific plasma waves that can accomplish the
speculative concepts put forth here.

This work was supported by DOE Contracts DE-FG02-06ER54851 and
DE-AC0276-CH03073.

\end{document}